\documentclass[12pt]{iopart}

\usepackage{graphicx}
\usepackage{color}
\usepackage{amssymb}
\usepackage{algorithmic}
\usepackage[boxed]{algorithm}

\begin{document}

\title{Discrete-time quantum walks generated by aperiodic fractal sequence of space coin operators}

\author{R. F. S. Andrade$^{1}$, and A. M. C. Souza$^{2,3}$}

\address{  $^{1}$ Instituto de F\'{i}sica, Universidade Federal da Bahia, 40210-340
Salvador Brazil.\\

$^{2}$Departamento de F\'{i}sica, Universidade Federal de
	Sergipe 49.100-000, S\~{a}o Cristov\~{a}o - Brazil. \\

   $^{3}$Department of Physics, University of Central Florida, Orlando,
FL 32816, USA.}

\ead{randrade@ufba.br}

\begin{abstract}
Properties of one dimensional discrete-time quantum walks are sensitive to the presence of inhomogeneities in the substrate, which can be generated by defining position dependent coin operators. Deterministic aperiodic sequences of two or more symbols provide ideal environments where these properties can be explored in a controlled way. This work discusses a two-coin model resulting from the construction rules that lead to the usual fractal Cantor set. Although the fraction of the less frequent coin $\rightarrow 0$ as the size of the chain is increased, it leaves peculiar properties in the walker dynamics. They are characterized by the wave function, from which results for the probability distribution and its variance, as well as the entanglement entropy were obtained. A number of results for different choices of the two coins are presented. The entanglement entropy has shown to be very sensitive to uncover subtle quantum effects present in the model.
\end{abstract}

\maketitle

\section{Introduction}

The discrete-time quantum walk (DTQW) on linear non-homogeneous
chains has attracted much attention recently. The related literature
presents several investigations on DTQW dynamics using chains where
the sequence of coin operations depends both in space and time.
Differently from the linear spread in time behavior of the wave
function for the homogeneous DTQW with just one coin, the
generalization based on time- and/or position-dependent coins
exhibits a much broader class of dynamic properties.

Consider DTQWs on chains with time-dependent coins where,
independently of the site in the chain, all coin operators have the
same action at a given time step. Here, the abundance of possible
different behavior is already observed when just two different
operators are present. For periodic sequences, the long time
behavior is still found to be ballistic. However, for coins selected
according to deterministically quasi-periodic sequences (e.g. the
two-coin Fibonacci sequence), the walk is characterized by a
sub-ballistic wave function spreading \cite{prl93,physa388}.
Sub-ballistic spreading is also obtained for L\'{e}vy two-coin
sequences \cite{pra76,pre76}, and the random choice of coin
operations leads to a diffusive behavior, similar to the  classical
random walk. In more general cases, we may consider a sequence of
different coins such that the number of different coins increases
with time \cite{pra73, pra80}. Given such a rich class of distinct
behavior, it is natural that specific selection can be made in such
a way to control the wave function spreading \cite{pra90}.

For position-dependent coins, where their action remains the same
for all time steps, the simplest cases is that in which only the
central coin is different from all the others \cite{qip8,qip9}.
Similarly, several works have discussed the dynamic properties of
chains with position-dependent coin operators based quasi-periodic
sequences like  Fibonacci, Thue-Morse, Rudin-Shapiro \cite{amb17},
or in which the coins are spatially inhomogeneous
\cite{pra80b,pre82}. More recently, an experimental apparatus
simulated a DTQW with phase position-dependent coins \cite{njp16}.
Finally, it is important to acknowledge that a thorough
identification of the properties of DTQW based on coins with both
time and position dependence is only starting. Early results have
already suggested it use as generator of probability distributions
\cite{pra95}.

Because of the quantum nature of the system, understanding the
entanglement properties of the DTQW wave functions is crucial in
many aspects, starting by identifying their differences with respect
to the classical walk. The entanglement entropy, which is an
important measure to characterize a large number of quantum systems,
has been also the main tool for the DTQW analysis
\cite{pra73b,pra81,qic11}. For instance, by considering
time-dependent coins it has been possible to show that disordered
coin sequences can maximize the entanglement \cite{prl111}.

In this work we present an analysis of the DTQW on a non-homogeneous
chain with space-dependent coin operators. The two-coin sequence is
generated by recursive use of the same geometrical rule that leads
to the Cantor set. However, instead of removing the sites placed in
the central segments in each iteration, these sites are assigned to
a the second type of coin operator, giving rise to the fractal
Cantor sequence. We present results for wave-function properties in
terms of the probability distribution as a function of space and
time, the standard deviation, and the entanglement entropy. The
dependence of the wave function properties on the choice of coins
($\theta_1$ and $\theta_2$) is illustrated by considering a number
of different angles defining the operators.

The rest of this work is organized as follows: In Sec. II we present
a brief review of the DTQW concepts and define the notation through
out the work.   Sec. III describes the construction of the Cantor
sequence. Results are discussed in Sec. IV, where we emphasize the
adequacy of the entanglement entropy to uncover details of the
quantum evolution of the system. Sec. V closes the work
with our concluding remarks.

\section{The DTQW framework}

Within the DTQW framework, the time evolution of a quantum particle
is described by the operator $\hat{W}$ that acts on the state vector
$|\Psi(x,t)\rangle=|\Phi(x)\rangle\otimes|\sigma(x,t)\rangle$, with
position $|\Phi (x) \rangle$ and coin $|\sigma(x,t)\rangle$
components. The discrete position and time variables are indicated,
respectively, by integer values of $x$ and $t\geq0$.

\begin{figure}
\centering
\includegraphics[width=8cm,angle=0]{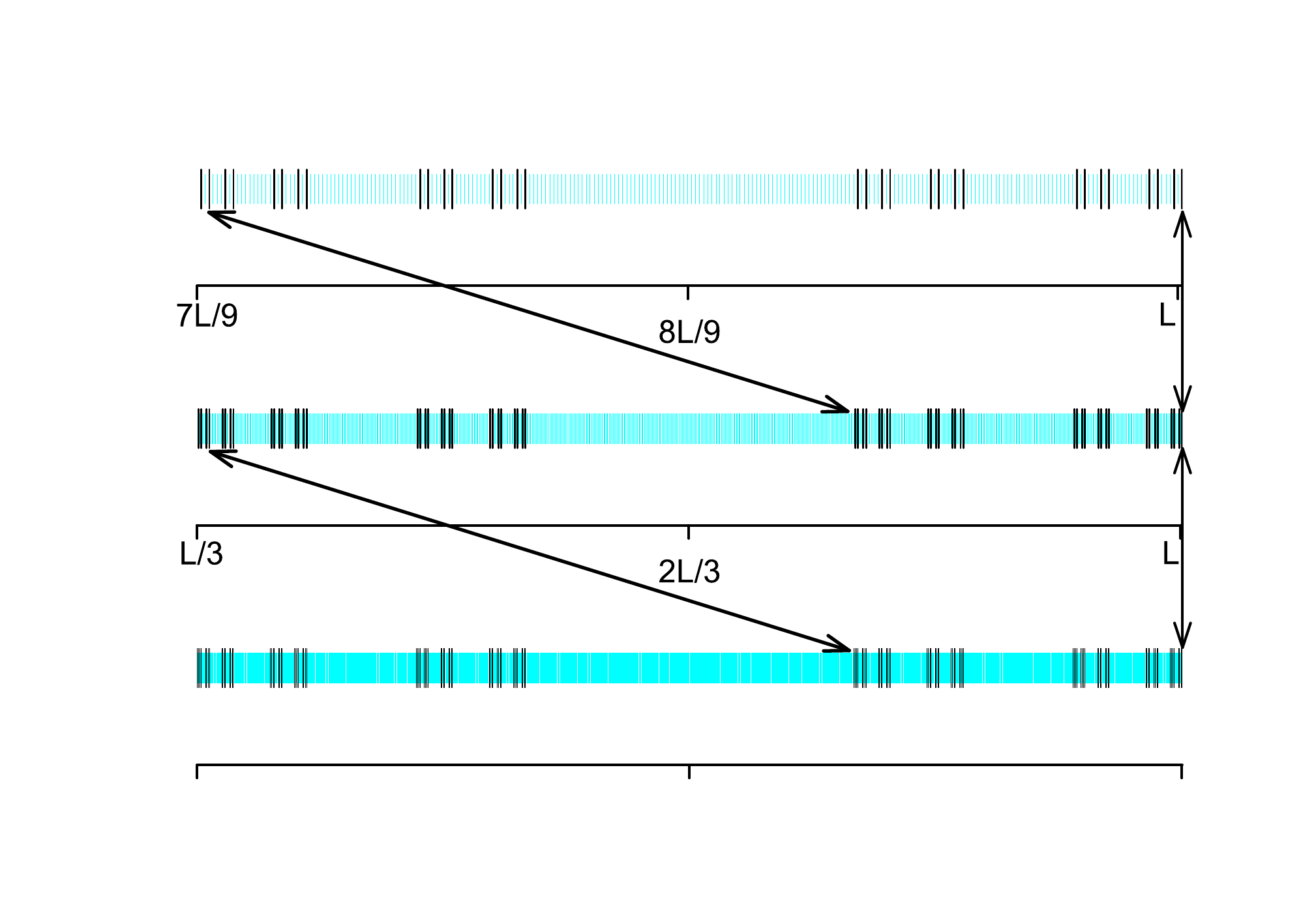}
 \caption{Three successive views of the fractal coin sequence from bottom to top. At each new view, a selected piece of the set corresponding to 1/3 of its displayed is enlarged by a factor 3. Black (cyan) vertical bars indicate the location of $\theta_1$ ($\theta_2$) coins.}
\label{fig1}
\end{figure}

At each time step $t$, the evolution is defined by the following unitary operations
\begin{equation}\label{eq1a}
    \hat{W}=\hat{S}(\hat{I}_{\infty}\otimes\hat{C}).
\end{equation}
\noindent Here, the coin operator $\hat{C}$ acts on the coin
components $|r\rangle$ and $|l\rangle$, and the shift operator
$\hat{S}$ updates the wave function magnitude at each chain sites
taking into account the coin state.

The coin operator $\hat{C}$ is expressed by the unitary matrix
\begin{equation}\label{eq2a}
    \hat{C}=\left(
  \begin{array}{cc}
    \cos \theta & \sin \theta  \\
    \sin \theta & -\cos \theta  \\
  \end{array}
\right),
\end{equation}
\noindent where $0 \le \theta \le \pi /2$. For non-homogeneous
space-dependent processes, $\theta=\theta(x)$. The above definition
of $\hat{C}$ is consistent with the expression of $\hat{S}$'s action
in terms of the coin degree states allowing the particle to move to
the right ($|r\rangle$) and to the left ($|l\rangle$) of $x$
\cite{Nayak}. Thus, the $\hat{S}$ operator can be expressed by

\begin{equation}\label{eq3b}
\hat{S} = \cos \theta \hat{S}_{r} |r \rangle\langle r|
    + \sin \theta \hat{S}_{l} |l\rangle\langle l|,
\end{equation}

\noindent where $x\in [-L,L]$ for a  chain of $N=2L+1$ sites,
$\hat{S}_{r}=\left[ \sum_{x=-L}^{L} |x+1\rangle \langle x| \right]$,
and $\hat{S}_{l}=\left[ \sum_{x=-L}^{L} |x-1\rangle\langle x|
\right]$.

Given that the wave function $|\Psi(x,t) \rangle$ entails all
properties of the walker, we study the physical properties of the
system by considering  $P(x,t)=| \langle x | \Psi(x,t) \rangle
|^{2}$, the probability distribution for finding the walker on the
site $x$ at time $t$.

In addition to $P(x,t)$, we also focus our attention on two further
functions derived from $|\Psi(x,t) \rangle$. The standard deviation
$\sigma (t) = \sqrt{\langle x(t)^{2} \rangle - \langle x(t)
\rangle^{2}}$, where $\langle x(t)^{m} \rangle = \sum_{x=-L}^{L}
x^{m} P(x,t)$, quantifies the wave function spreading. Next, to
quantify the quantum entanglement present in the pure state $|
\Psi(x,t) \rangle$, we consider the entanglement entropy defined as
$S_{E} (t) = - Tr(\hat{\rho}_c log_{2} \hat{\rho}_{c})$. Here,
$\hat{\rho}_{c}(t)$ is the reduced density operator given by
$\hat{\rho}_{c}(t)= Tr_{x} \hat{\rho}(x,t)$, where $\hat{\rho}(x,t)
= | \Psi(x,t) \rangle \langle \Psi(x,t) |$ is the density operator
of the total system, and the partial trace $Tr_{x}$ is taken over
all sites in the chain.

In order to separate the contribution of the two coin states, the
wave function $|\Psi(x,t)\rangle$ can be described using two
component vector amplitudes as \cite{Nayak,pra81,nos13}
\begin{equation}\label{psidec}
    | \Psi (x,t) \rangle = \left( \psi_{r} (x,t), \psi_{l}(x,t) \right)^{\textbf{T}} .
\end{equation}
\noindent Here, $\psi_{r} (x,t)$ and $\psi_{l} (x,t)$  are,
respectively, the amplitudes of a walker with $|r\rangle$ and
$|l\rangle$ coin internal degrees at site position $x$ and time $t$,
while $*^{\textbf{T}}$ represents a transposed matrix.

Using this decomposition, the probability distribution can be expressed by
\begin{equation}
P(x,t) = |\psi_{r} (x,t)|^{2} + |\psi_{l} (x,t)|^{2}
\end{equation}
while the entanglement entropy becomes \cite{pra73b}
\begin{equation} \label{ee1}
S_{E}(t) = - p \log_{2} (p) - (1-p) \log_{2} (1-p).
\end{equation}
\noindent Here, $p=(1+\sqrt{1-4(AC-|B|^{2})})/2 $, $A=\sum_{x} |\psi_{r} (x,t)|^{2}$, $B=\sum_{x} \psi_{r} (x,t) \psi_{l}^{*} (x,t) $ and $C= \sum_{x} |\psi_{l} (x,t)|^{2}$.

\section{The Cantor coin sequence}

As mentioned Sec. I, the coin operator $\hat{C}$ is position
dependent according to the Cantor sequence, which is built in a
iterative way in a close relation to the procedure leading to the
Cantor set. However, instead of carrying out the deletion of $1/3$
of the continuous intervals in the previous generations, as in the
usual Cantor set building procedure, we start at generation $g=0$
with a single type-1 coin associated with $\theta_1$. At $g=1$, this
coin is replaced by a (type-1,type-2,type-1) three-coin sequence,
where the type-2 coin is associated with $\theta_2$. All subsequent
$g>1$ generations are obtained from the previous one by preforming
the replacement operations
$\theta_{1}\rightarrow\theta_{1}\theta_{2}\theta_{1}$ and
$\theta_{2}\rightarrow\theta_{2}\theta_{2}\theta_{2}$. The $g$th
generation of the coin sequence has $3^g$ sites and coins, and
actual fractal is obtained in the $g\rightarrow\infty$ limit. The
first generations of the Cantor sequence are represented as
$\theta_{1} \rightarrow \theta_{1}\theta_{2}\theta_{1} \rightarrow
\theta_{1}\theta_{2}\theta_{1}\theta_{2}\theta_{2}\theta_{2}\theta_{1}\theta_{2}\theta_{1}
\rightarrow ...$ .

At a given generation $g$, the obtained structure consists of
isolated $\theta_1$ coins, with left and right neighborhoods formed
by sequences of $3^\ell$ $\theta_2$ coins, with $0\leq\ell<g$. The
number of sites of type-1 coins is $N_{g,1}=2^g$, and the
corresponding set has a fractal dimension $d_f=\ln(2)/\ln(3)$, while
the set of type-2 coins has $d_f=1$, consistent  with $N_{g,2}/N
\rightarrow 1$ when $g \rightarrow \infty$. This stays in contrast
to the behavior of other quoted sequences (e.g. Fibonacci), where
the fraction of sites associated to a given coin operator does not
vanish in the $g \rightarrow \infty$ limit. To obtain and compare
results for complete sequences at each generation, the values of $L$
are such that $N_g=3^g=2L_g+1$. As we discuss in the next section,
the properties of the DTQW in the Cantor chain are actually distinct
in many aspects from those reported in previous investigations.

\begin{figure}
\centering
\includegraphics[width=8cm,angle=0]{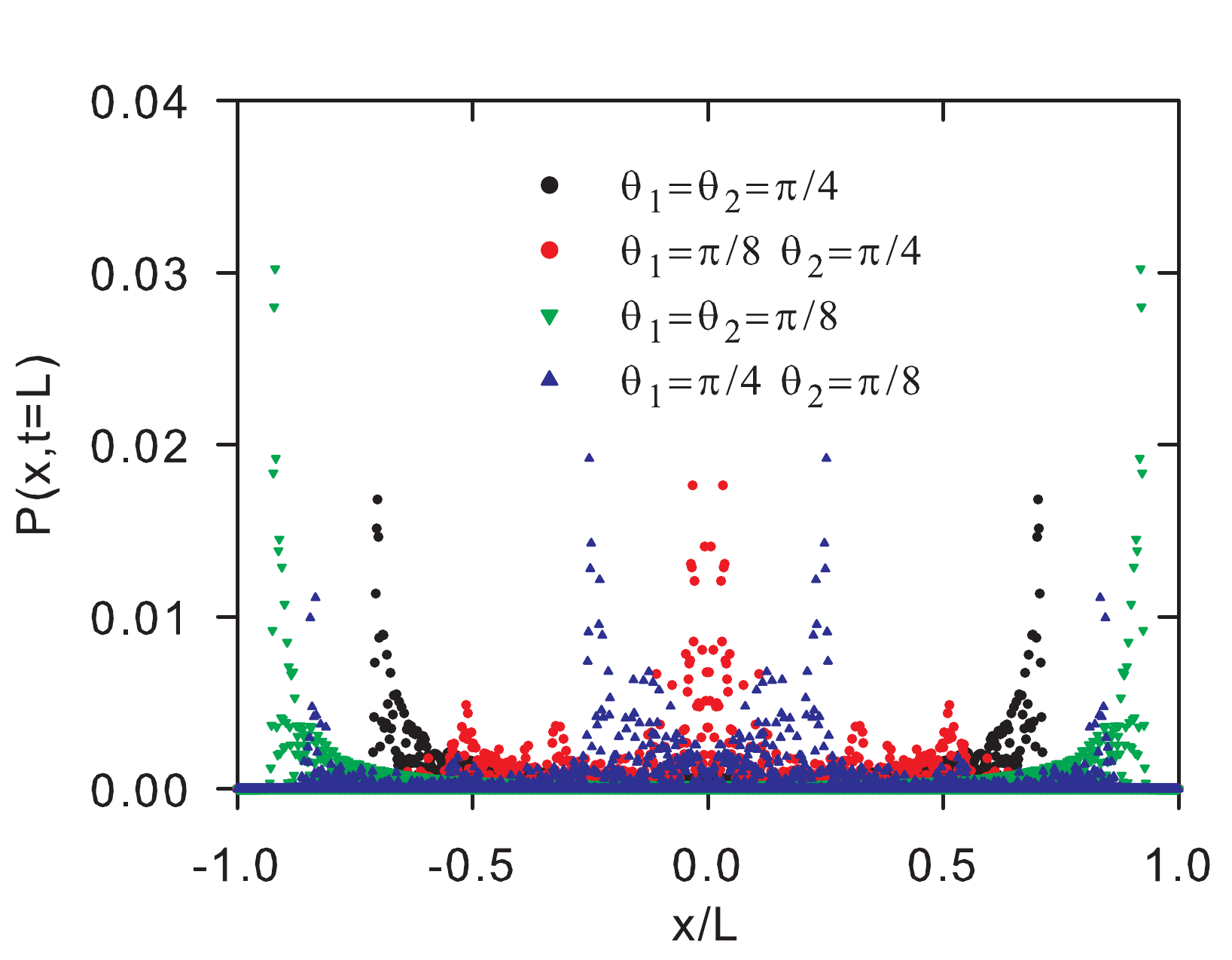}
\caption{Probability distribution $P(x,t=L)$ as a function of $x$ for $N=2187$ coins using a Cantor sequence. In panel (a), we consider equal angles cases $\theta_{1}=\theta_{2}=\pi/4$ (black circles) and $\theta_{1}=\theta_{2}=\pi/8$ (red squares). In panel (b), $\theta_{1}=\pi/8$ and $\theta_{2}=\pi/4$ (black circles), and $\theta_{1}=\pi/4$ and $\theta_{2}=\pi/8$ (red squares).}
\label{fig2}
\end{figure}

\begin{figure}
\centering
\includegraphics[width=8cm,angle=0]{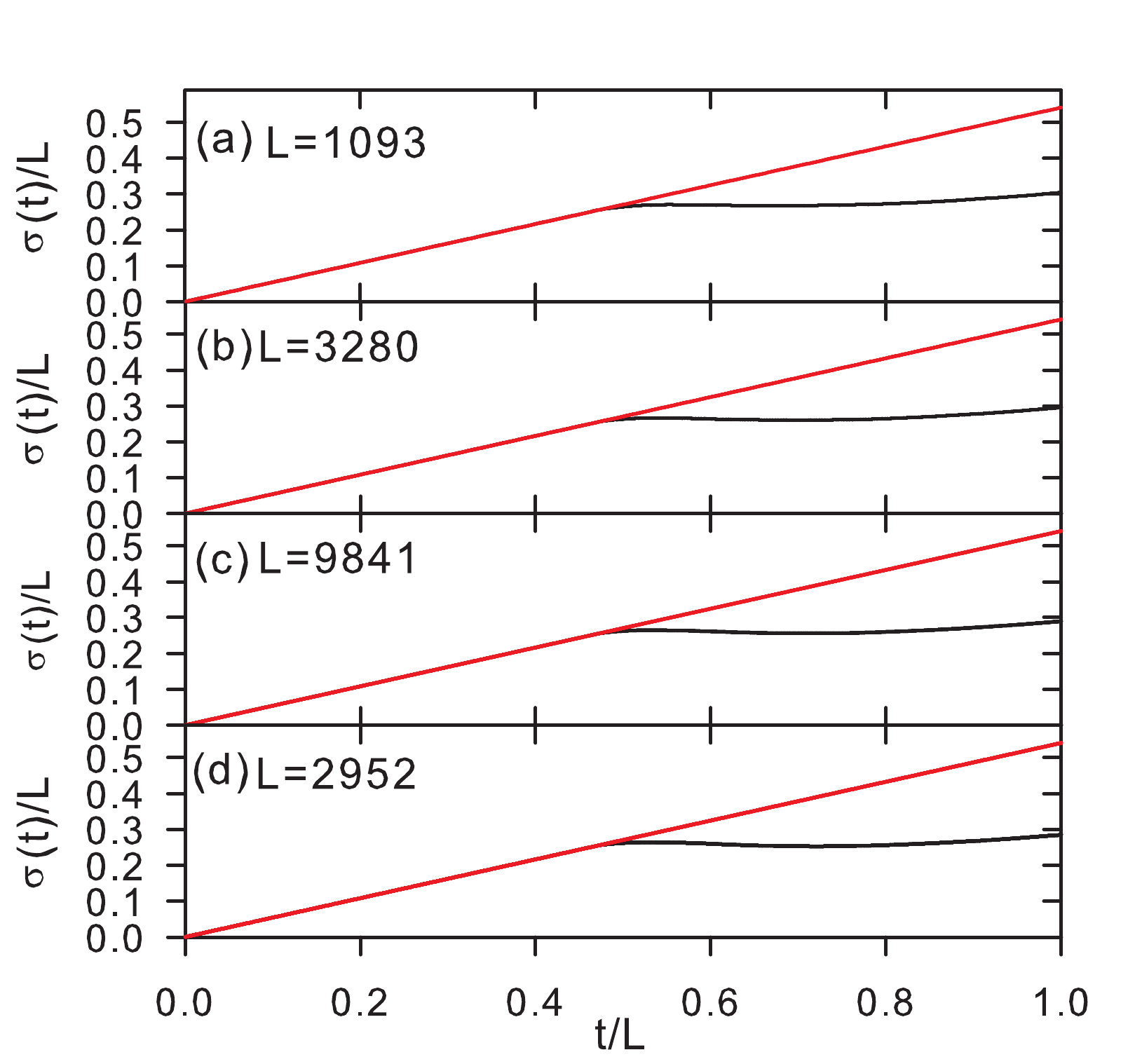}
\caption{Standard deviation $\sigma(t)/L$ as a function of $t/L$ for
(a) $L=1093$, (b) $L=3280$, (c) $L=9841$ and (d) $L=29524$.
Red (gray) and black lines correspond, respectively, to $\theta_{1}=\theta_{2}=\pi/4$ ,
and $\theta_{1}=\pi/8$ and $\theta_{2}=\pi/4$. The two curves are coincident for $t/L \le 0.4694$.}
\label{fig3}
\end{figure}

\begin{figure}
\centering
\includegraphics[width=8cm,angle=0]{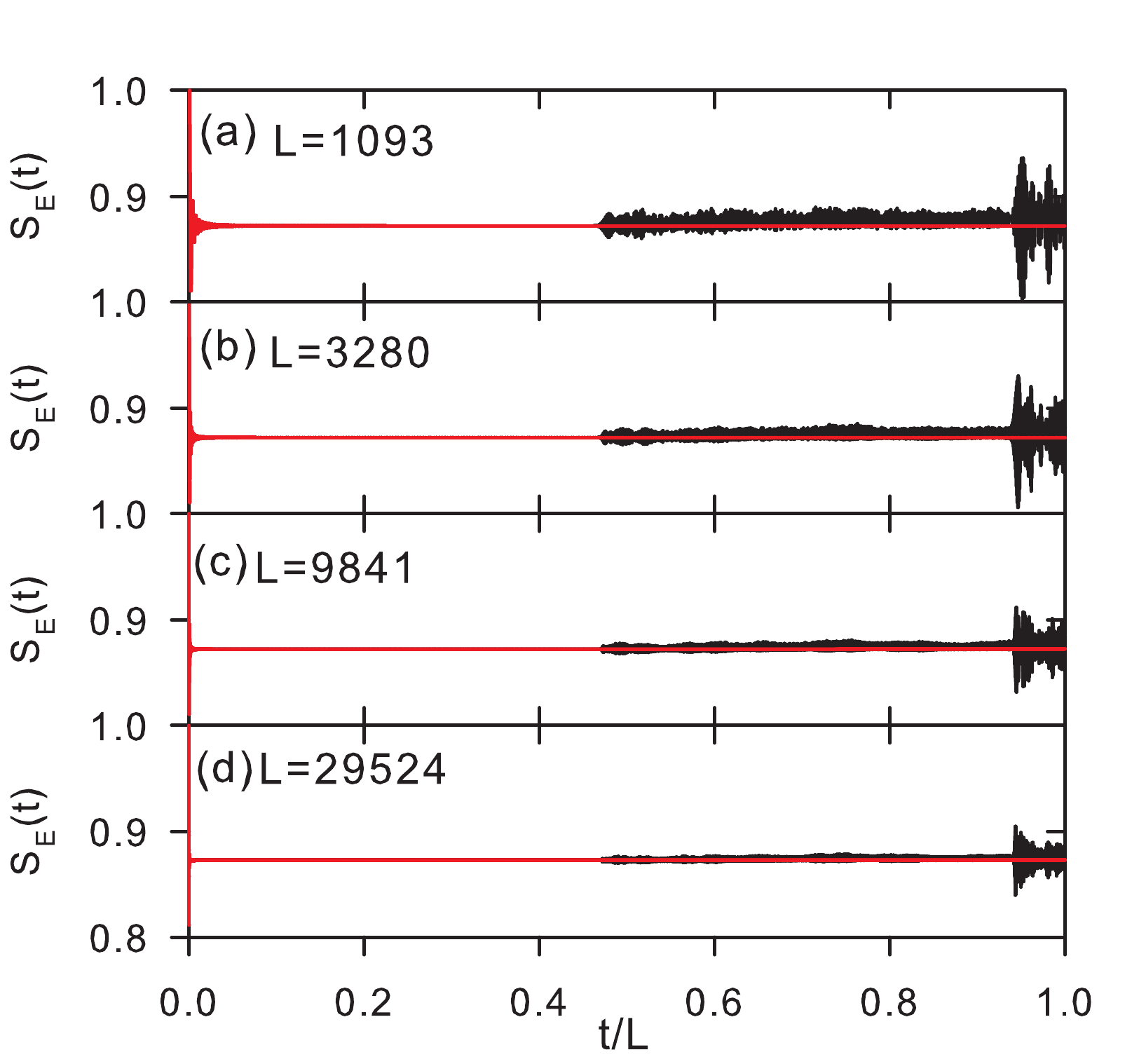}
\caption{$S_{E}(t)$ as a function of  $t/L$ for (a) $L=1093$, (b)
$L=3280$, (c) $L=9841$ and (d) $L=29524$. Red (gray) line and black
correspond, respectively, to $\theta_{1}=\theta_{2}=\pi/4$, and
$\theta_{1}=\pi/8$ and $\theta_{2}=\pi/4$. The two curves are
coincident for $t/L \le 0.4694$.}
\label{fig4}
\end{figure}

Figure \ref{fig1} represents the coin sequence at the $8$th
generation ($N_g=6561, L_g=3280$). Black (cyan) symbols represent the
site positions corresponding to coin $\theta_{1}$ ($\theta_{2}$).
The third part of the set is zoomed-in twice to illustrate the
fractal structure of the sequence.

\section{Results}
We start with the analysis of the influence of different coins in
the behavior of $P(x,t)$ illustrated in Figure \ref{fig2}. We
consider $N=2187$ ($L=1093$), and data corresponds to $t=L=1093$.
Once the initial condition is always $|\Psi(x,t=0)>=|x=0>\otimes
\frac{1}{\sqrt(2)}(|r>+i|l>) $, choosing $t=L$ has the advantage of
providing information on $P(x,t)$ in a condition where
breaking of the fractality in the coin sequence caused by finite-size lattices is relatively small.

For the sake of comparison to the uniform walk, we draw results for $\theta_1=\theta_2$ (a) and for $\theta_1\ne\theta_2$ (b). For equal coins, the results indicate that the greatest probabilities of finding the walker occur at the edges of the chain, which is consistent with known ballistic spread. The difference between the position of the peaks
is due only to the effective size of each leap, which is controlled
by $\cos \theta_2$. For different coins, the peaks of $P(x,t)$ no
longer stay at the chain edges, but have moved to more internal
sites of the chain. This leads to a decrease in the spread of the
walker. It can be observed that, by switching $\theta_1
\leftrightarrow \theta_2$, the resulting patterns are different due
to the different effective leap size associated to each coin by its
angle, and the number of places where they are present. Other signatures for this same behavior become explicit in plots for $\sigma(t)$ and $S_E(t)$. Such an effect on dynamics of the system is somewhat expected, given the
large difference in the values of $N_1$ and $N_2$, as well as the
particular spatial distribution of the two coin operators.

\begin{figure}
\centering
\includegraphics[width=8cm,angle=0]{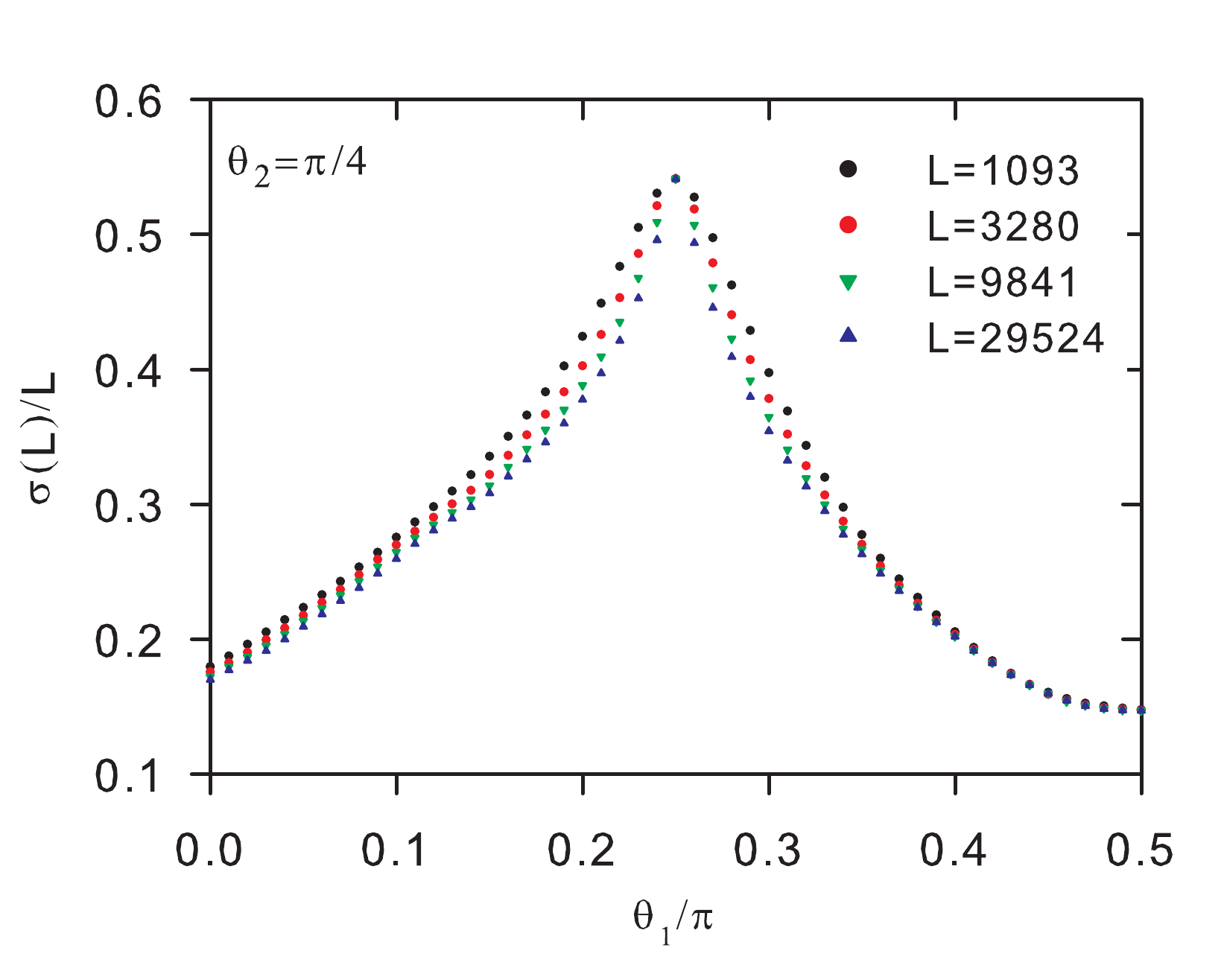}
\caption{Standard deviation $\sigma(t=L)/L$ as a function of $\theta_{1}$ for
$\theta_{2}=\pi/4$ and $L=1093$, $3280$, $9841$ and $29524$.} \label{fig5}
\end{figure}

\begin{figure}
\centering
\includegraphics[width=8cm,angle=0]{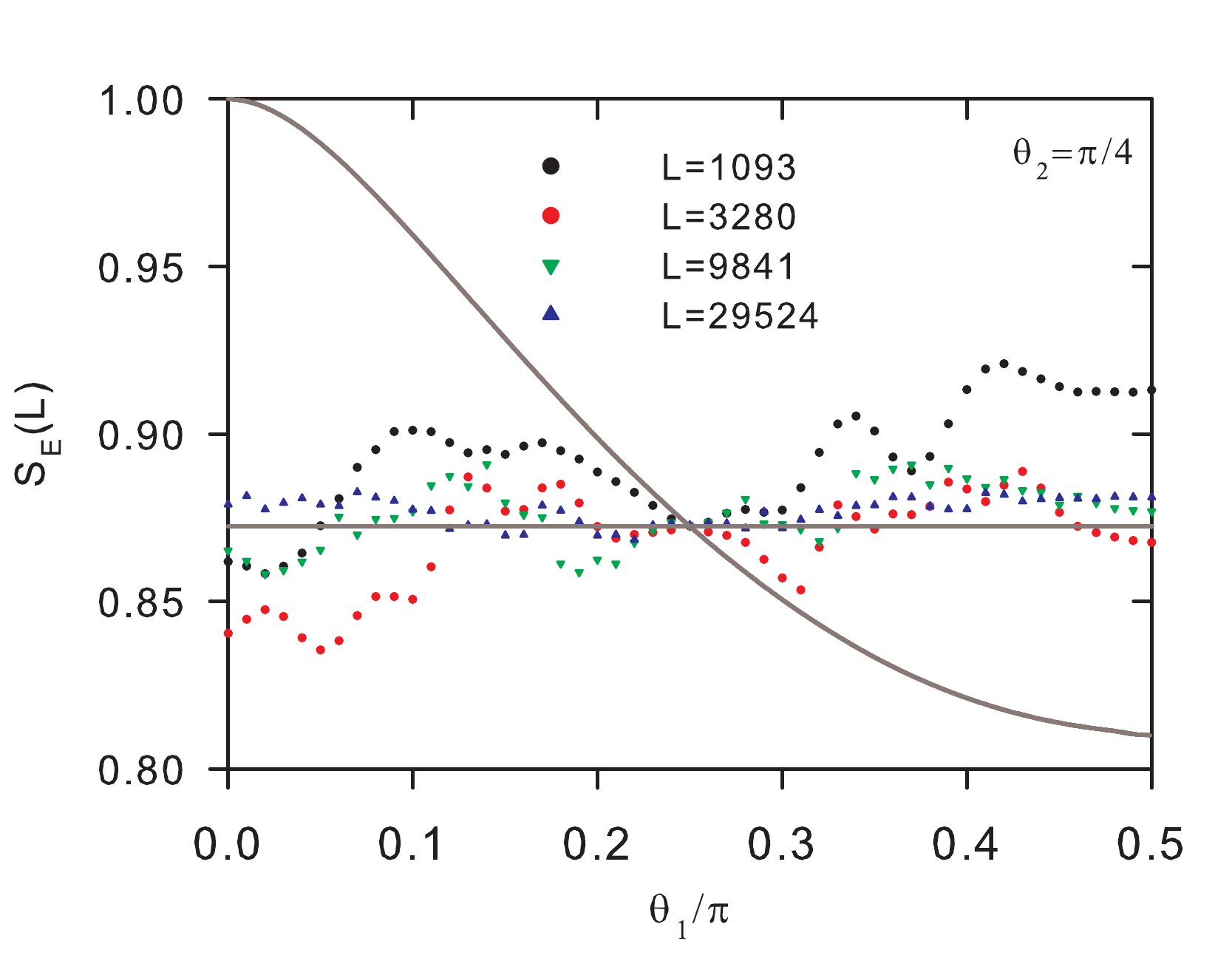}
 \caption{Entanglement entropy $S_{E}(t=L)$ as a function of $\theta_{1}$
 for $\theta_{2}=\pi/4$ and $L=1093$, $3280$, $9841$ and $29524$.
Gray curve decreasing represents the case of one coin
($\theta_{2}=\theta_{1}$) and the gray curve constant
($S_{E}=0.8724$) the case $\theta_{2}=\theta_{1}=\pi/4$.}
\label{fig6}
\end{figure}

\begin{figure}
\centering
\includegraphics[width=8cm,angle=0]{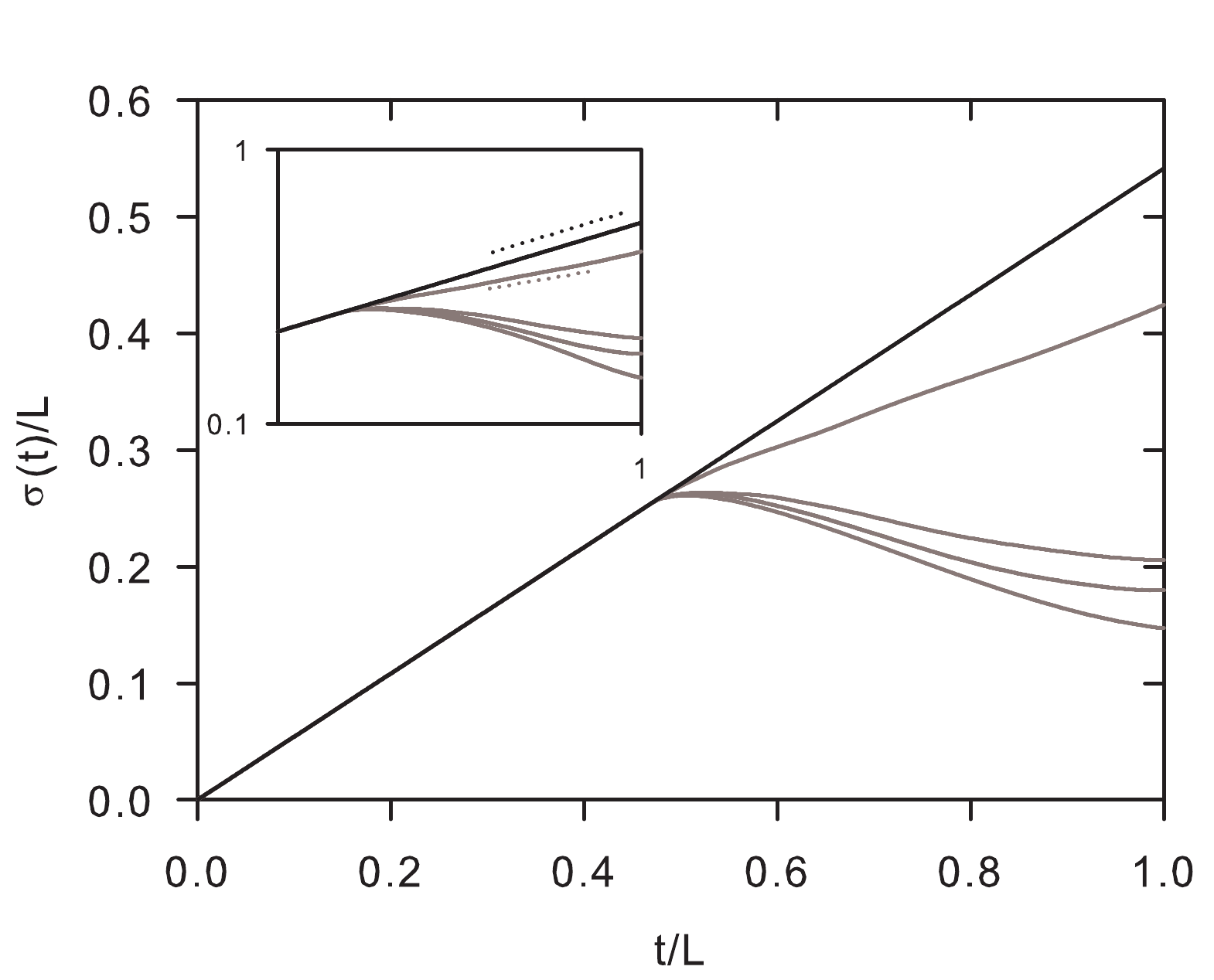}
\caption{Standard deviation $\sigma(t)/L$ [$L=1093$] as a function of time for $\theta_{2}=\pi/4$ and $\theta_{1}=\pi/4$ (black curve) and $\theta_{1}=4\pi/5$, $2\pi/5$, $\pi$, $8\pi/5$ (gray curves from above to below). In the inset we draw double logarithmic scale plots. Dot line shows the dependence $\sigma(t) \propto t$.}
\label{fig7}
\end{figure}

As the Cantor sequence only reaches an exact fractal structure in
the $g\rightarrow\infty$ limit, one might expect that finite-size
effects caused by the finite chain impact the obtained
results. Figure \ref{fig3} shows the time evolution of the standard
deviation $\sigma (t)$ for successive chain sizes [(a) $L=1093$, (b)
$L=3280$, (c) $L=9841$ and (d) $L=29524$]. Each panel contains two
curves: the red line, which corresponds to
$\theta_{1}=\theta_{2}=\pi/4$, has been inserted for the purpose of
comparison. The black line refers to $\theta_{1}=\pi/8$ and
$\theta_{2}=\pi/4$. It is characterized by linear growth that is
coincident with the behavior for $\theta_{1}=\theta_{2}$ until a
value $t_c$ where it deviates from the straight line. The results do
not show significative dependence of $\sigma(t)$ on $L$, suggesting
absence of the influence of the chain size.

On the other hand, Figure \ref{fig4} shows that the chain size
influences the the time evolution of the entanglement entropy
$S_{E}(t)$. In the four panels, which correspond to the same chain
sizes used in Figure \ref{fig3}, the curve for
$\theta_{1}=\theta_{2}=\pi/4$ and  $2\theta_{1}=\theta_{2}=\pi/4$
again overlap until $t=t_c$. At this value, $S_{E}(t)$ for
$2\theta_{1}\neq \theta_{2}$ is no longer constant, as it happens
for the condition $\theta_{1}=\theta_{2}$. However, $S_{E}(t>t_c)$
shows considerable dependence on the chain size $L$, in opposition
to $L$ independent pattern shown in Figure \ref{fig3}.

These results can be better appreciated when we analyze the behavior
of the two functions at fixed $t=L$, for different values of
$\theta_{1}$ and fixed $\theta_{2}=\pi/4$, as illustrated in Figures
\ref{fig5} and \ref{fig6}. Again, we can see that the effect of
finite lattice size on $\sigma$ is very small, characterized by a
monotonic decreasing behavior of $\sigma(t=L)$ as a function of $L$
 $\forall \theta_1$, while $S_{E}$ shows a strong size-dependent
pattern.

The lines in Figure \ref{fig6} have different meanings. The
sinusoidal dashed curve represents $S_{E}(t=L)$ when
$\theta_{1}=\theta_{2}$ while the constant straight line indicates
$S_{E}=0.8724$, which is obtained for $\theta_{1}=\theta_{2}=\pi/4$.
The numerical results suggest that $S_{E}$ converges to this value
in the $L=t \rightarrow \infty$ limit, independently of
$\theta_{1}$. Such a convergence, if happens to be confirmed, has a
highly non uniform character.

\begin{figure}
\centering
\includegraphics[width=8cm,angle=0]{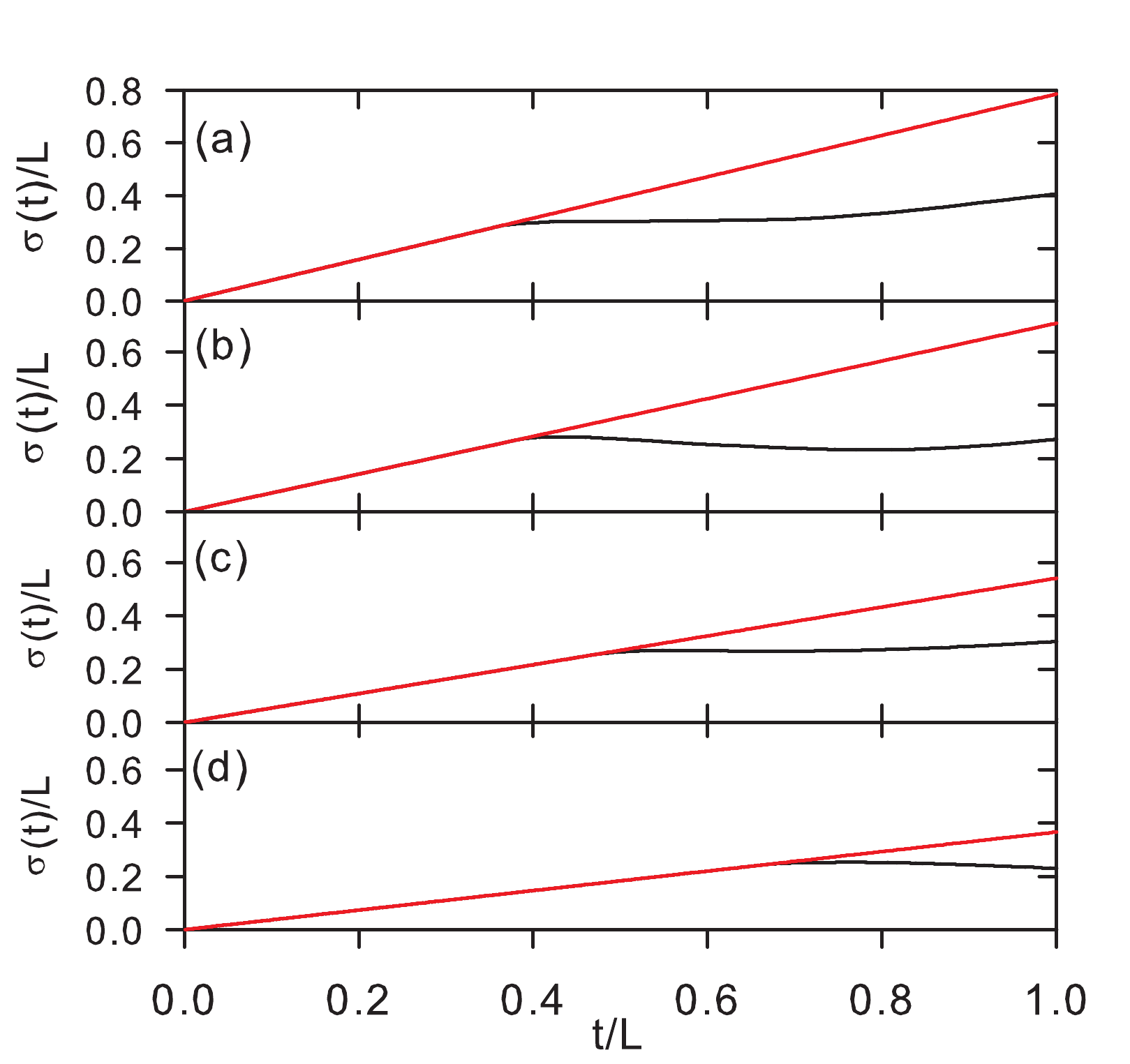}
\caption{Standard deviation $\sigma(t)/L$ versus time for $\theta_{1}=\theta_{2}=\pi/8$ and $\theta_{1}=\pi/4$ and $\theta_{2}=\pi/8$ (a), $\theta_{1}=\theta_{2}=\pi/6$ and $\theta_{1}=\pi/3$ and $\theta_{2}=\pi/6$ (b), $\theta_{1}=\theta_{2}=\pi/4$ and $\theta_{1}=\pi/8$ and $\theta_{2}=\pi/4$ (c),  $\theta_{1}=\theta_{2}=\pi/3$ and $\theta_{1}=\pi/6$ and $\theta_{2}=\pi/3$ (d). In each panel, the curve for $\theta_{1}=\theta_{2}$ (red) is drawn to to emphasize the difference between the ballistic and the the complex two-coin regime for $\theta_{1} \neq \theta_{2}$ (black). $L=1093$}
\label{fig8}
\end{figure}

\begin{figure}
\centering
\includegraphics[width=8cm,angle=0]{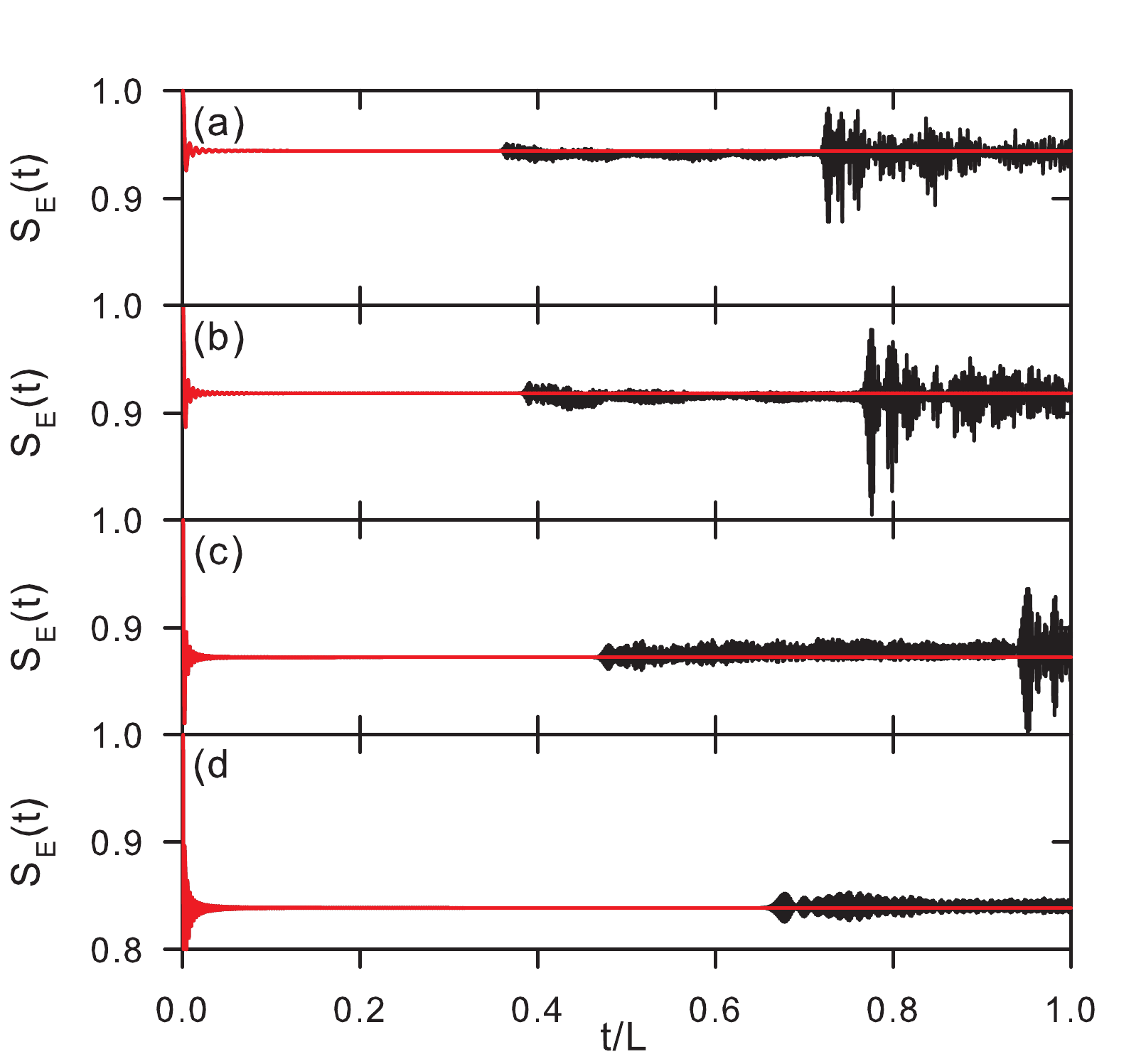}
\caption{Entanglement entropy $S_{E}(t)$ versus time $t$ for (a)
$\theta_{1}=\theta_{2}=\pi/8$ and $\theta_{1}=\pi/4$ and $\theta_{2}=\pi/8$.
(b) $\theta_{1}=\theta_{2}=\pi/6$ and $\theta_{1}=\pi/3$ and $\theta_{2}=\pi/6$;
(c) $\theta_{1}=\theta_{2}=\pi/4$ and $\theta_{1}=\pi/8$ and $\theta_{2}=\pi/4$;
(d) $\theta_{1}=\theta_{2}=\pi/3$ and $\theta_{1}=\pi/6$ and $\theta_{2}=\pi/3$;
Red and black lines indicate, respectively, $\theta_{1}=\theta_{2}$ and $\theta_{1} \neq \theta_{2}$.
$L=1093$.}
\label{fig9}
\end{figure}

It is interesting to observe that the value of $t_c$ where the
behavior of $\sigma(t)$ and $S_{E}(t)$ deviate from the ballistic
regime does not depend on $\theta_1\neq\theta_2$, as illustrated in
Figure \ref{fig7}.

\begin{figure}
\centering
\includegraphics[width=8cm,angle=0]{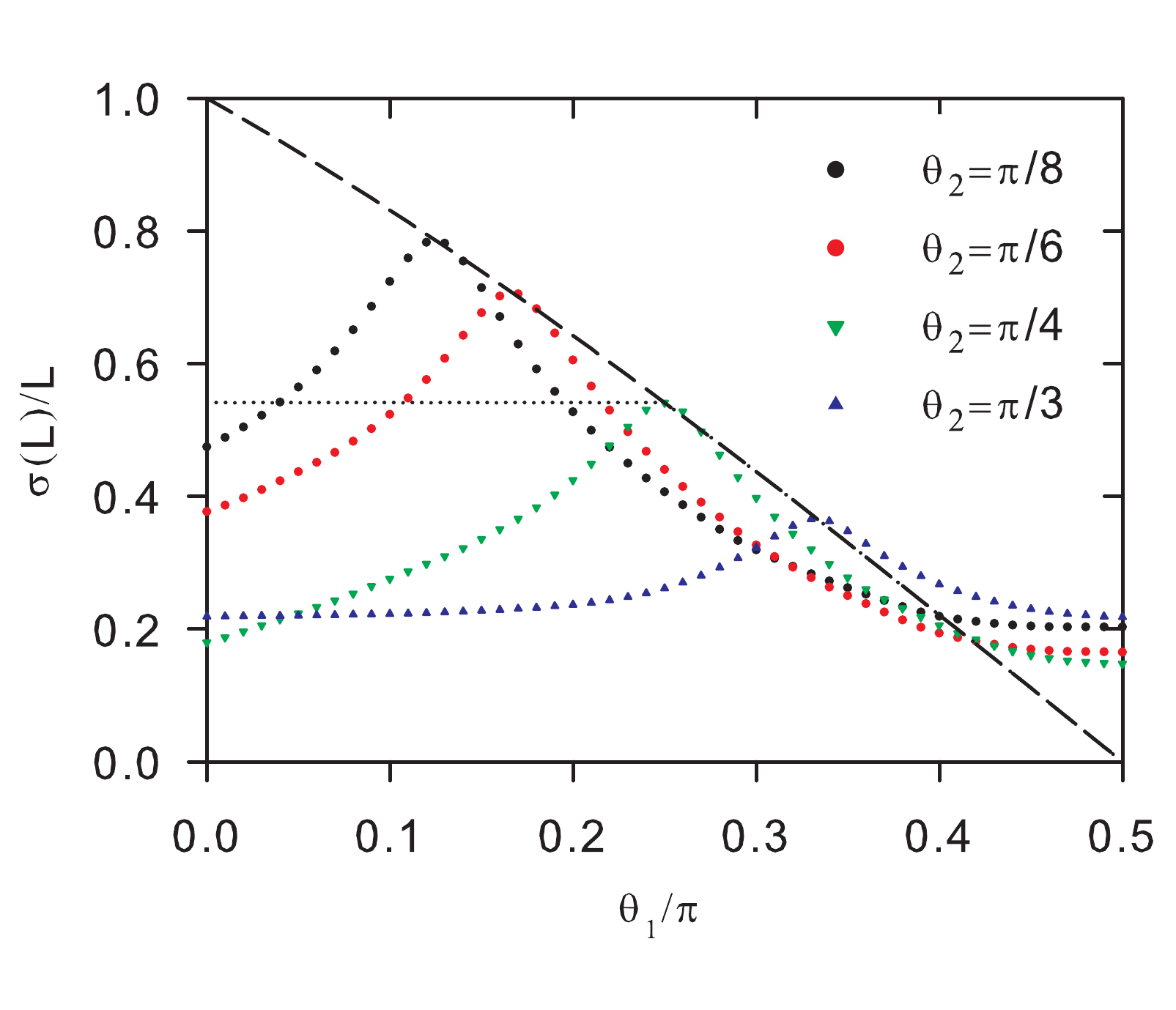}
\caption{Standard deviation $\sigma(t=L)$ versus $\theta_{1}$ for $\theta_{2}=\pi/8$ (black), $\theta_{2}=\pi/6$ (red), $\theta_{2}=\pi/4$  (green), $\theta_{2}=\arccos(5/9)$ (blue), $\theta_{2}=\pi/3$ (cyan). Dashed curve represents the case of the unique coin $\theta_{1}=\theta_{2}$ and dotted curve the case of the two periodic coins with $\theta_{2}=\pi/4$. $L=1093$. }
\label{fig10}
\end{figure}

To illustrate the influence of $\theta_{2}$ on our results, let us
consider now a chain fixing $L=1093$. Figure \ref{fig8} shows the
time evolution of $\sigma(t)$ for $\theta_{2}=\pi/8, \pi/6, \pi/4,$
and $\pi/3$. Once we have shown in Figure \ref{fig7} that the value
of $t_c$ is independent of $\theta_1$, we limit ourselves to show
only two curves in each panel, much as done in Figure \ref{fig3}.
The first curve shows the ballistic behavior at
$\theta_{1}=\theta_{2}$, while the second curve corresponds either
to $\theta_{1}=2\theta_{2}$ or $\theta_{1}=\theta_{2}/2$.

The behavior of $\sigma(t)$ in all panels is very similar to those
shown in Figure \ref{fig3}: when $\theta_{1} \ne \theta_{2}$ the
walker keeps the ballistic behavior until $t_{c}(\theta_{2})$, when it leaves this regime. The dependence of $t_{c}$ on $(\theta_{2})$ is clearly observed by observing the panels, which
also illustrate that the velocity of the ballistic spreading
decreases with $\theta_2$.

Figure \ref{fig9} shows the behavior of $S_{E}(t)$ for the same
conditions used in Figure \ref{fig7}. The first three panels also
make evident that the value of $t > t_c$ where the fluctuation in
$S_{E}(t)$ undergo a significative amplitude increase coincide with
$t\simeq 2t_c$. On the other hand, once $t_c(\theta_2=\pi/3)>0.5$,
the second transition to the large amplitude regime fails to be
observed.

\begin{figure}
\centering
\includegraphics[width=8cm,angle=0]{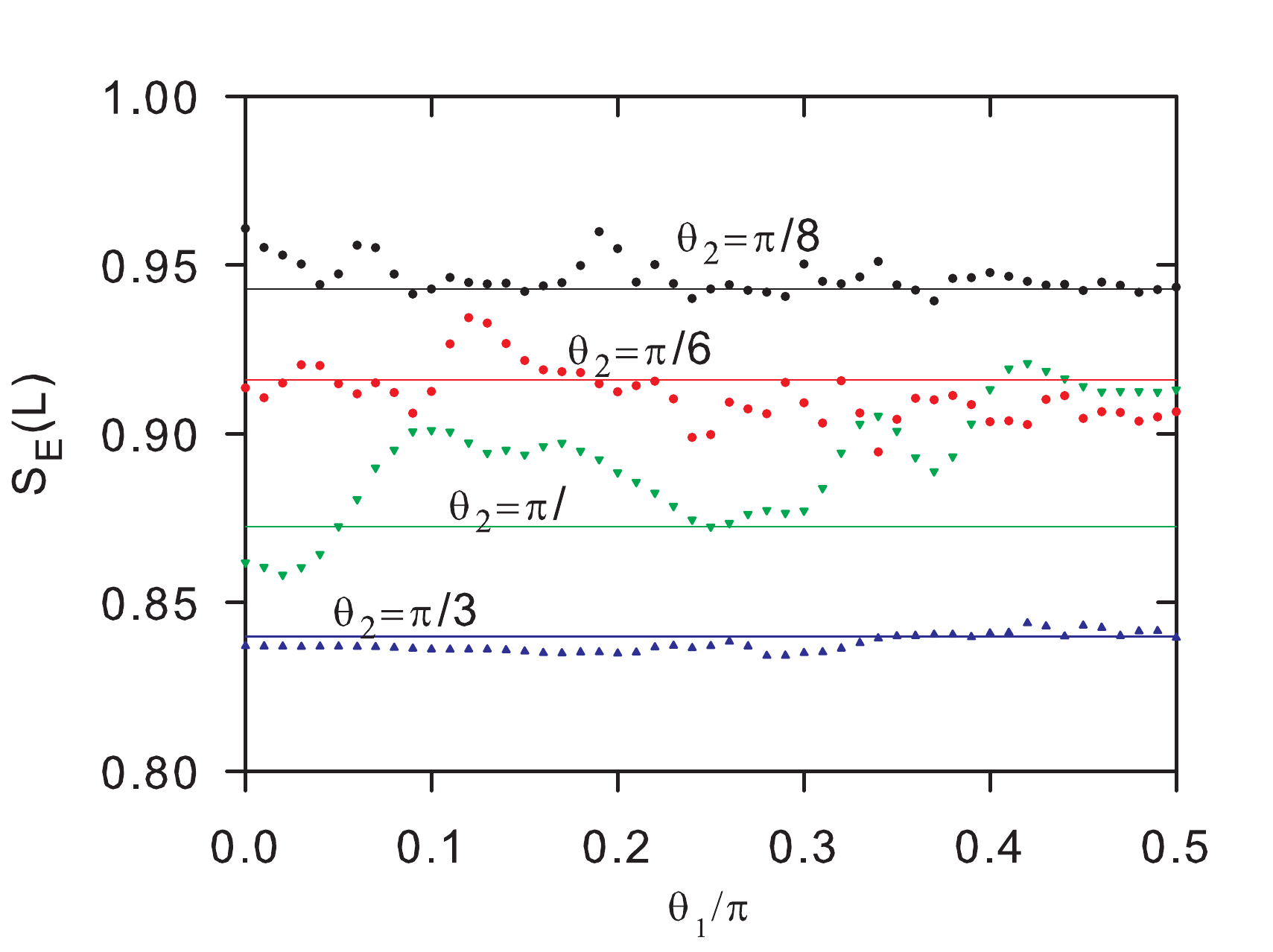}
\caption{Entanglement entropy $S_{E}(t=L)$ versus
$\theta_{1}$ for $\theta_{2}=\pi/8$ (black data),
$\theta_{2}=\pi/6$ (red data), $\theta_{2}=\pi/4$  (green data),
$\theta_{2}=\pi/3$ (blue data). $L=1093$.} \label{fig11}
\end{figure}

The results for $\sigma(t=L)$ and $S_{E}(t=L)$ as a
function of  $\theta_{1}$ for different values of $\theta_{2}$ are
shown in Figs. \ref{fig10} and \ref{fig11}. They follow the same
features displayed, respectively, in Figures \ref{fig5} and
\ref{fig6}. We observe that, $\forall \theta_2 $, the peaks of $\sigma(t=L)$,  occur at the locus of the corresponding curve when $\theta_1=\theta_2$ (dashed line). The results also indicate the existence of a value $\theta_2^*\simeq \arccos (5/9)$ at which the derivative of $\sigma(t=L)$ also coincides with the derivative of the same curve for $\theta_1=\theta_2$. Regarding the behavior of $S_{E}(t=L)$, Fig. \ref{fig11} shows that it fluctuates about the time
independent value of $S_{E}(t)$ when $\theta_1=\theta_2$. As the
time independent value of $S_{E}(t)$ for the uniform chain does
depend on $\theta_2$, the four $S_{E}(t=L)$ curves as a function of
$\theta_1$ oscillate around $\theta_2$-dependent values.

All the discussed features of $\sigma$ and $S_{E}$ reflect in some
way the complexity of DTQW dynamics induced by the Cantor sequence.
Some of them can be understood, at least at a qualitative level, by
the structure of the chain itself. Let us first remark that, after
leaving the origin at $t=0$, the walker has its motion influenced
only by the constant angle $\theta(x) = \theta_{2}$ during the first
$L/3$ sites either to left or to the right. Thus, the value of
$\theta_{1}$ plays no role in the walker's motion as long as
$t/L<1/3$, so that the ballistic spread is an expected result.

Next, we observe that the walker's forward velocity is limited by
the term $\cos \theta_2$ in the diagonal elements of Eq.
(\ref{eq2a}). This effect is also present in the pure ballistic
spread when $\theta_1 = \theta_{2}$, as indicated by the behavior of
$P(x,t=L)$ shown in Figure \ref{fig2}, as well as from the different
slopes of the straight in the lines for $\sigma$ as a function of
$t$ in Figure \ref{fig8}.

\begin{figure}
\centering
\includegraphics[width=8cm,angle=0]{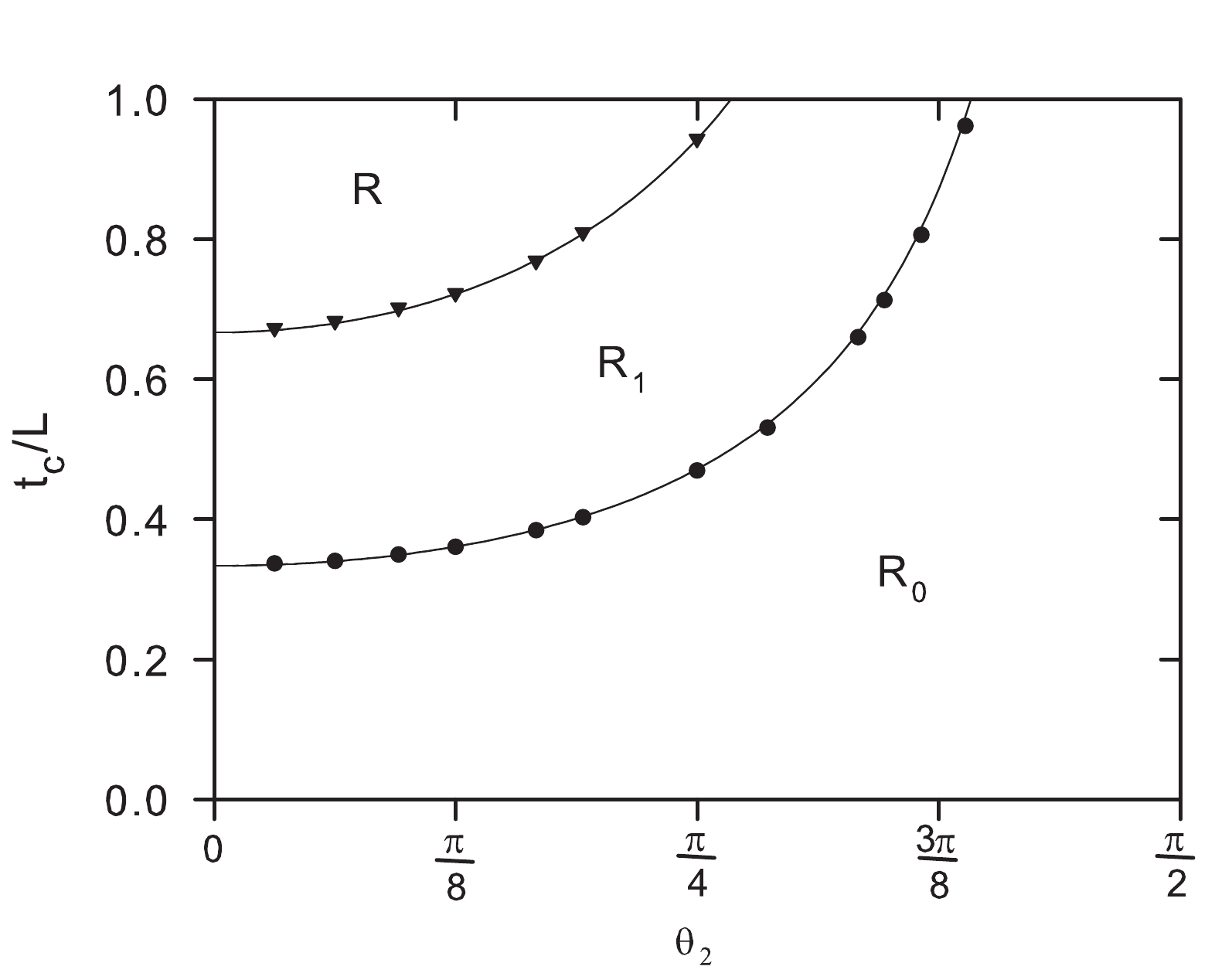}
\caption{Diagram showing the regions where the walker dynamics
undergo transitions in the behavior of $S_{E}$. R$_0$: Not influenced
by the value of $\theta_{1}$; R$_1$: Influenced by the value of
$\theta_{1}$ for $t>t_{c}$; R$_2$: Influenced by the value of
$\theta_{1}$ for $t>t_{c}$ with the emergence of a large amplitude
pattern at $t=2t_{c}$. Circles indicate critical time in which the
difference between $\sigma(t)$ for $\theta_{1} \neq \theta_{2}$ and
$\sigma(t)$ for $\theta_{1} = \theta_{2}$ is $0.01\%$. It coincides
with the emergence of small amplitude modulation in $S_{E}$.
Triangles indicate the value of $t$ where $S_{E}$ starts developing
large amplitude pattern.
The lines indicate the value of $t_c$ and $2t_c$ given by Eq. (\ref{eqtc}).} \label{fig12}
\end{figure}

Thus, it follows that the walker starts to feel the influence of
$\theta_{1}$ operators only after
\begin{equation}\label{eqtc}
t_{c}=L/(3\cos\theta_{2})
\end{equation}
\noindent time steps. The above expression is an  increasing
function of $\theta_{2}$, consistent with the results in Figures
\ref{fig8} and \ref{fig9}. For the purpose of comparing the above
expression with the numerical results of the integration of
evolution equation, we defined $t_{c}$ as the value of $t$ at which
the difference between $\sigma(t)$ for $\theta_{1} \neq \theta_{2}$
and $\sigma(t)$ for $\theta_{1} = \theta_{2}$ reaches $0.01\%$.

A comparison between the results for $t_{c}$ as a function of
$\theta_{2}$ obtained by Eq. (\ref{eqtc}) and the numerical results
is presented in Figure \ref{fig12}. It shows that $t_{c}$ increases
from $t_{c} = L/3$ for $\theta_{2}=0$ to $t_{c}=L$ for $\theta_{2}
\simeq 1.23$. When $\theta_{2} > 1.23$ the walker never leaves the ballistic  regime when $t\le L$ and, as a consequence, the fractal coin sequence requires longer time interval in order to
start influencing the walker's dynamics. The Figure \ref{fig12} also
shows a comparison between the values of $2t_{c}$ as a function of
$\theta_{2}$ with the values of $t$ where the second transition to
large amplitude fluctuation of $S_E$ are observed.

The amplitude of oscillations for $S_{E}(t>t_c)$ starts with a
rather smooth sinusoidal dependency, but it rapidly develops a
complex shape. For $\theta_{2}=\pi/3$ it is still possible to
identify the superposition of effects due to short and long
frequencies in the pattern. However, the overall behavior becomes
soon very complicated. The same observation applies to the sudden
increase in the amplitude of oscillations when $t>2t_{c}$.

\begin{figure}
\centering
\includegraphics[width=8cm,angle=0]{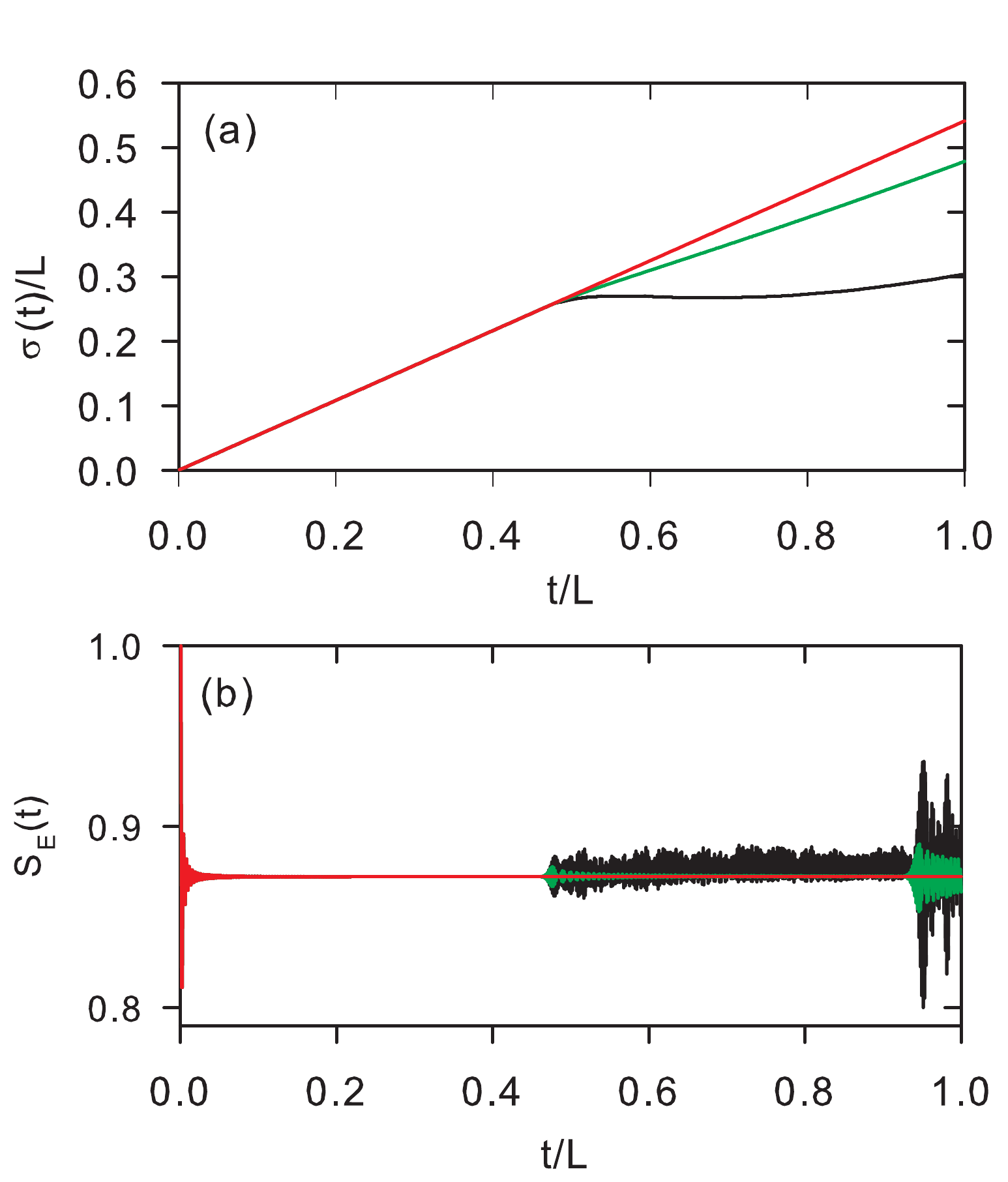}
\caption{Comparison of the standard deviation $\sigma(t)$ and entanglement entropy $S_{E}(t)$ versus $t$, when $\theta_{1}=\pi /8$ and $\theta_{2}=\pi /4$, for the two-scatter (green) and Cantor sequence (black) models. Curves in red correspond to the homogeneous model $\theta_{1}=\theta_{2}=\pi /4$. $L=1093$.}
\label{fig13}
\end{figure}

Finally, for the purpose of obtaining a qualitative measure of the influence of the the full fractal structure of the Cantor sequence in our results, we also evaluated the behavior of a much simpler two-scatter model. It consists of an open linear chain with the same length as the Cantor sequence model, where $\theta_1$ coins are assigned to all but the sites located at $x=\pm L/3$, where the dynamics is described by a $\theta_2$ coin. The results for $\theta_1=\pi /8$ and $\theta_2=\pi /4$ are displayed in Fig. \ref{fig13}. It becomes clear that the value of $t_c$ for the Cantor sequence is the result that is mostly influenced by the two first $\theta_2$ sites, while all other features show strong differences. The time evolution of $\sigma(t)$ shows a much stronger depart from the ballistic dynamics for the Cantor sequence model. Along the same line, the behavior of $S_E(t)$ for the two-scatter model is relatively short lived for $t>t_c$ and has an almost sinusoidal pattern as compared become much complex behavior for the Cantor sequence model. Regarding the second transition to larger magnitude values for $S_E(t)$, we also notice the same short lived sinusoidal pattern, as well as a slight difference in the values of $t$ where it occurs for the two models.

\section{Conclusions}
We have analyzed the DTQW on a non-homogeneous one-dimensional substrate formed by an aperiodic set of $\theta_{1}$ coins immersed on a much larger number of  $\theta_{2}$ coins. $\theta_{1}$ coins are placed according to the rules of construction of the Cantor set. The walks were constrained to a time interval chosen as to avoid the effect of reflections on the boundaries of the substrate. Our numerical results for the probability distribution, for the spread of the wave packet and for the entanglement entropy show that this
intertwined two-coin distribution leaves a very peculiar signature on the walker dynamics. Scale invariance is achieved already for a small number of generations in the construction of the sequence. The entanglement entropy has shown to be a very sensitive measure of quantum properties of the dynamics, revealing changes in the walkers dynamics that are left unnoticed by other measures. However, despite the great sensitivity by the presence of coin $\theta_{1}$ found in the entropy entanglement for finite systems, when the system tends to an infinite size, this influence must disappear, remaining only the signature in standard deviation $\sigma$.

We explored several results for different values of the phase angles in the $\theta_{1}$ and $\theta_{2}$ coins. They indicate that $\theta_2$ plays a major role in the quantum behavior. The comparison with the results for $\theta_{1}=\theta_{2}$ indicates a completely different pattern as soon as we let $\theta_{1} \neq \theta_{2}$. This emerging pattern remains relatively unchanged for all values $\theta_{1}$.

For walks starting at center of the chain, the effect of $\theta_{1}$ appears only after the necessary number of steps $t_{c}$ for the walker to reach the first $\theta_{1}$ coin. This event immediately causes changes in the behavior of $\sigma$ and $S_{E}$.

Unlike the case of two coins in a periodic sequence, in which the walker has a ballistic diffusion, or in the cases of aperiodic sequences, in which we have sub-ballistic dynamics \cite{prl93,physa388}, the Cantor fractal sequence with two coins has a ballistic regime for $t <t_c$ and switches to a more complex dynamics where its value can even decrease for some limited time intervals. The comparison with the two-scatter model leads to the understanding of the transition times but also emphasizes the richness of the considered model. Given the recent advances in producing DTQW experiments with with phase position-dependent coins, we hope the interesting effects reported in this work can be better understood.

\section{Acknowledgements}
The authors acknowledge the financial support of Brazilian agency CNPq. Both authors benefit from the support of the Instituto Nacional de Ci\^{e}ncia e Tecnologia para Sistemas Complexos (INCT-SC).



\begin{thebibliography}{40}

\bibitem{prl93} P. Ribeiro, P. Milman, and R. Mosseri, Aperiodic Quantum Random Walks, Phys. Rev. Lett. \textbf{93},
190503 (2004).

\bibitem{physa388} A. Romanelli, The Fibonacci quantum walk and its classical trace map, Physica A \textbf{388}, 3985 (2009).

\bibitem{pra76} A. Romanelli, Measurements in the Levy quantum walk, Phys. Rev. A \textbf{76}, 054306 (2007).

\bibitem{pre76} A. Romanelli, R. Siri, and V. Micenmache, Sub-ballistic behavior in quantum systems with Levy noise, Phys. Rev. E \textbf{76}, 037202 (2007).

\bibitem{pra73} M. C. Banuls, C. Navarrete, A. Perez, Eugenio Roldan, and J. C.
Soriano, Quantum walk with a time-dependent coin, Phys. Rev. A
\textbf{73}, 062304 (2006).

\bibitem{pra80} A. Romanelli, Driving quantum-walk spreading with the coin operator, Phys. Rev. A \textbf{80}, 042332 (2009).

\bibitem{pra90} M. Montero, Invariance in quantum walks with time-dependent coin
operators, Phys. Rev. A \textbf{90}, 062312 (2014).

\bibitem{qip8} N. Konno, One-dimensional discrete-time quantum walks on random environments, Quant. Inf. Proc. \textbf{8}, 387 (2009).

\bibitem{qip9} N. Konno, Localization of an inhomogeneous discrete-time quantum walk on the line, Quant. Inf. Proc. \textbf{9}, 405 (2010).

\bibitem{amb17} C. V. Ambarish, N. Lo Gullo, Th. Busch, L. Dell'Anna, and C. M.
Chandrashekar, Dynamics and energy spectra of aperiodic
discrete-time quantum walks, Phys. Rev. E \textbf{96}, 012111
(2017).

\bibitem{pra80b} N. Linden, and J. Sharam, Inhomogeneous quantum walks, Phys. Rev. A \textbf{80}, 052327 (2009).

\bibitem{pre82} Y. Shikano, and H. Katsura, Localization and fractality in inhomogeneous quantum walks with self-duality, Phys. Rev. E \textbf{82}, 031122 (2010).

\bibitem{njp16} P. Xue, H. Qin, B. Tang, and B. C. Sanders, Observation of quasiperiodic dynamics in a onedimensional
quantum walk of single photons in space, N. J. Phys. \textbf{16},
053009 (2014).

\bibitem{pra95} M. Montero, Quantum and random walks as universal generators of probability distributions, Phys. Rev. A \textbf{95},
062326 (2017).

\bibitem{pra73b} G. Abal, R. Siri, A. Romanelli, and R. Donangelo, Quantum walk on the line: Entanglement and nonlocal initial conditions, Phys. Rev. A \textbf{73}, 042302, 069905(E) (2006).

\bibitem{pra81} A. Romanelli, Distribution of chirality in the quantum walk: Markov process and
entanglement, Phys. Rev. A \textbf{81}, 062349 (2010).

\bibitem{qic11} Y. Ide, N. Konno, and T. Machida, Entanglement for discrete-time quantum walks on the line, Quant. Inf. Comput. \textbf{11}, 855 (2011).

\bibitem{prl111} R. Vieira, E. P. M. Amorim, and G. Rigolin, Dynamically Disordered Quantum Walk
as a Maximal Entanglement Generator, Phys. Rev. Lett. \textbf{111},
180503 (2013).

\bibitem{Nayak} A. Nayak and A. Vishwanath, Quantum walk on the Line, quant-ph/0010117 (2000).

\bibitem{nos13}  A. M. C. Souza, and R. F. S. Andrade, Coin state properties in quantum walks, Sci. Rep. \textbf{3}, 1976 (2013).

\end{thebibliography}
\end{document}